# Comment on "Amplified emission and lasing in photonic time crystals"

Jagang Park[1,†], Hee Chul Park[2], Kyungmin Lee[3], Jonghwa Shin[4], Jung-Wan Ryu[2], Wonju Jeon[1], Namkyoo Park[5], and Bumki Min[3,*]

[1]*Department of Mechanical Engineering, Korea Advanced Institute of Science and Technology, Daejeon 34141, Republic of Korea*

[2]*Center for Theoretical Physics of Complex Systems, Institute for Basic Science (IBS) Daejeon 34126, Republic of Korea*

[3]*Department of Physics, Korea Advanced Institute of Science and Technology, Daejeon 34141, Republic of Korea*

[4]*Department of Material Sciences and Engineering, Korea Advanced Institute of Science and Technology, Daejeon 34141, Republic of Korea*

[5]*Department of Electrical and Computer Engineering, Seoul National University, Seoul 08826, Republic of Korea*

[*]Corresponding author. Email: bmin@kaist.ac.kr

[†]Present address: *Department of Electrical Engineering and Computer Sciences, University of California, Berkeley, CA 94720, USA*

**Lyubarov *et al.* (Research Articles, 22 July 2022, p. 425) claim that the spontaneous emission rate of an atom vanishes at the momentum gap edges of photonic Floquet media. We show that their theoretical prediction is based on assumptions that result in misleading interpretations on the spontaneous emission rate in photonic Floquet media.**

One of the most notable features of photonic Floquet media (or photonic *temporal* crystals) is the presence of momentum gaps (MGs), inside which the complex-conjugated eigenfrequencies of photonic Floquet states govern the propagation, amplification, and parametric oscillation and lasing (*1–7*). Recently, Lyubarov *et al.* claim that the density of photonic states, into which an excited atom can radiate, decreases down to zero at the MG edge so that the spontaneous emission is prohibited (*8*). Here, we point out that their analysis does not properly consider the non-Hermiticity of the photonic Floquet media and equivalently the non-orthogonality of the photonic Floquet modes, which lead to an incorrect estimation on the spontaneous emission rate (SER). The implication is substantial because the correct calculation when considering the non-Hermiticity reveals that the SER is non-zero at the MG edge via the Petermann factor (PF).

The non-Hermiticity of the photonic Floquet medium can be explicitly revealed by examining the following Floquet Hamiltonian matrix (*4*);

$$H_F = \begin{bmatrix} \ddots & & \vdots & & \udots \\ & \widetilde{H}^0 + \Omega \mathbf{I} & \widetilde{H}^{-1} & \widetilde{H}^{-2} & \\ \cdots & \widetilde{H}^{+1} & \widetilde{H}^0 & \widetilde{H}^{-1} & \cdots \\ & \widetilde{H}^{+2} & \widetilde{H}^{+1} & \widetilde{H}^0 - \Omega \mathbf{I} & \\ \udots & & \vdots & & \ddots \end{bmatrix},$$

where the 2-by-2 matrices $\widetilde{H}^m = (1/T) \int_0^T dt \, \exp(im\Omega t) H(t)$ are the Fourier components of the time-periodic *effective* Hamiltonian $H(t)$ derived from the Maxwell's equations and their first-order approximation in the driving strength is given in (*4*). The off-diagonal block matrices

($\widetilde{H}^m$, $m \neq 0$) describe the coupling between the original bands ($\widetilde{H}^0$) and their shifted replicas ($\widetilde{H}^0 + m\Omega\mathbf{I}$). The Hermiticity is apparently broken due to the off-diagonal elements in off-diagonal block matrices (i.e., $\left(\widetilde{H}^m\right)^\dagger \neq \widetilde{H}^{-m}$ when $m \neq 0$); here one can see that the non-Hermiticity arises due to the coupling between positive- (negative-) frequency bands and replicated negative- (positive-) frequency bands (*4, 9*).

To prove that the non-Hermiticity gives rise to the non-orthogonality of the Floquet modes, we calculated the Floquet right and left eigenvectors of a photonic Floquet medium along with the complex-valued band structure (Fig. 1A and B), based on the eigenvalue problems for the Floquet Hamiltonian matrix: $(H_F - \omega_\alpha^m \mathbf{I})|F_{R\alpha}^m\rangle = 0$ and $\langle F_{L\alpha}^m|(H_F - \omega_\alpha^m \mathbf{I}) = 0$ (*10*). While the calculated band structure turns out to be identical to that shown in (*8*), evaluation of the inner product of two Floquet right eigenvectors relevant to the formation of the MG, $\langle F_{R+}^0|F_{R-}^1\rangle$, clarifies that they are not orthogonal to each other and become coalesced at the MG edge (see Fig.1C). The PF, which is a measure of non-orthogonality, is defined as follows;

$$\mathrm{PF} = \frac{\langle F_{R+}^0|F_{R+}^0\rangle\langle F_{L+}^0|F_{L+}^0\rangle}{|\langle F_{L+}^0|F_{R+}^0\rangle|^2}.$$

In Fig. 1D, the *diverging PF at the MG edges* is clearly seen, from which we can conclude that the assumption made by Lyubarov *et al.* on the orthogonality of the Floquet modes (see the Supplementary Materials of (*8*)) needs to be carefully examined.

Lyubarov *et al.* stated that the rate of spontaneous emission vanishes because the density of states, $\rho \propto k^2(\partial\omega/\partial k)^{-1}$, goes down to zero at the MG edges. However, *the vertical slope of the dispersion relation ($\partial\omega/\partial k \to \infty$) at the MG edges does not guarantee a vanishing SER in non-Hermitian systems*. The momentum-gap-like band structures with a vertical slope can be observed in other non-Hermitian photonic systems (e.g., photonic crystals with radiative loss

(*11*, *12*)) and, for these types of systems, it has been proven theoretically (*12*) and verified experimentally (*13*) that the SER is enhanced at the MG edge. In fact, the diverging PF comes into play in these cases, making the SER at the MG edge rather enhanced than that in the band (*12*, *13*). These theoretical prediction and experimental observation contradict evidently the above argument by Lyubarov *et al.*

To summarize, the prediction of a vanishing SER at the MG edge is shown to be incorrect because the non-Hermiticity of photonic Floquet media and the non-orthogonality of the Floquet modes were not properly considered in (*8*). We hope that further discussion follows about the correct estimation of the SER at the MG edges of photonic Floquet media.

# References

1. E. S. Cassedy, Dispersion relations in time-space periodic media: Part II - Unstable interactions. *Proc. IEEE*. **55**, 1154–1168 (1967).

2. J. R. Reyes-Ayona, P. Halevi, Observation of genuine wave vector (k or β) gap in a dynamic transmission line and temporal photonic crystals. *Appl. Phys. Lett.* **107** (2015), doi:10.1063/1.4928659.

3. N. Chamanara, Z. L. Deck-Léger, C. Caloz, D. Kalluri, Unusual electromagnetic modes in space-time-modulated dispersion-engineered media. *Phys. Rev. A*. **97** (2018), doi:10.1103/PhysRevA.97.063829.

4. N. Wang, Z. Q. Zhang, C. T. Chan, Photonic Floquet media with a complex time-periodic permittivity. *Phys. Rev. B*. **98** (2018), doi:10.1103/PhysRevB.98.085142.

5. E. Galiffi, P. A. Huidobro, J. B. Pendry, Broadband nonreciprocal amplification in luminal metamaterials. *Phys. Rev. Lett.* **123** (2019), doi:10.1103/PhysRevLett.123.206101.

6. J. Park, B. Min, Spatiotemporal plane wave expansion method for arbitrary space–time periodic photonic media. *Opt. Lett.* **46**, 484–487 (2021).

7. S. Lee, J. Park, H. Cho, Y. Wang, B. Kim, C. Daraio, B. Min, Parametric oscillation of electromagnetic waves in momentum band gaps of a spatiotemporal crystal. *Photonics Res.* **9**, 142 (2021).

8. M. Lyubarov, Y. Lumer, A. Dikopoltsev, E. Lustig, Y. Sharabi, M. Segev, Amplified emission and lasing in photonic time crystals. *Science* **377**, 425–428 (2022).

9. J. Park, H. Cho, S. Lee, K. Lee, K. Lee, H. C. Park, J.-W. Ryu, N. Park, S. Jeon, B. Min, Revealing non-Hermitian band structure of photonic Floquet media. *Sci. Adv.* **8**, eabo6220 (2022).

10. M. V Berry, Mode degeneracies and the petermann excess-noise factor for unstable lasers. *J. Mod. Opt.* **50**, 63–81 (2003).

11. B. Zhen, C. W. Hsu, Y. Igarashi, L. Lu, I. Kaminer, A. Pick, S. L. Chua, J. D. Joannopoulos, M. Soljačić, Spawning rings of exceptional points out of Dirac cones. *Nature*. **525**, 354–358 (2015).

12. A. Pick, B. Zhen, O. D. Miller, C. W. Hsu, F. Hernandez, A. W. Rodriguez, M. Soljačić, S. G. Johnson, General theory of spontaneous emission near exceptional points. *Opt. Express*. **25**, 12325 (2017).

13. L. Ferrier, P. Bouteyre, A. Pick, S. Cueff, N. H. M. Dang, C. Diederichs, A. Belarouci, T. Benyattou, J. X. Zhao, R. Su, J. Xing, Q. Xiong, H. S. Nguyen, Unveiling the Enhancement of Spontaneous Emission at Exceptional Points. *Phys. Rev. Lett.* **129**, 83602 (2022).

**Acknowledgments**

We thank Prof. Hansuek Lee, Prof. Fabian Rotermund and Prof. Yong-Hee Lee for helpful discussions.

**Funding:** This work was supported by National Research Foundation of Korea (NRF) through the government of Korea (NRF-2021R1C1C100631612 and 2017R1A2B3012364) and Institute of Information & communications Technology Planning and Evaluation (IITP) grant funded by the Korea government (MSIT) (No. 2022-0-00624). The work was also supported by the center for Advanced Meta-Materials (CAMM) funded by Korea Government (MSIP) as Global Frontier Project (NRF-2014M3A6B3063709).

**Competing interests:** The authors declare that they have no competing interests.

**Figure captions**

**Figure 1 Non-Hermiticity of photonic Floquet media.** (a) Real and (b) imaginary part of the non-Hermitian Floquet band structure. All the assumptions are identical to those used in the paper by Lyubarov et al. (c) Magnitude of the inner product of two different right Floquet eigenvectors, $|\langle F_{R-}^1|F_{R+}^0\rangle|$, relevant to the opening of the MG. Unity magnitude values at the MG edges signify coalescing of two right Floquet eigenvectors. (d) Petermann factor (PF) calculated in the first temporal Brillouin zone. The PF diverges at the MG edges, which are the second-order non-Hermitian degeneracies, i.e., exceptional points. The insets show enlarged views of PF near the MG edges.

# Figures

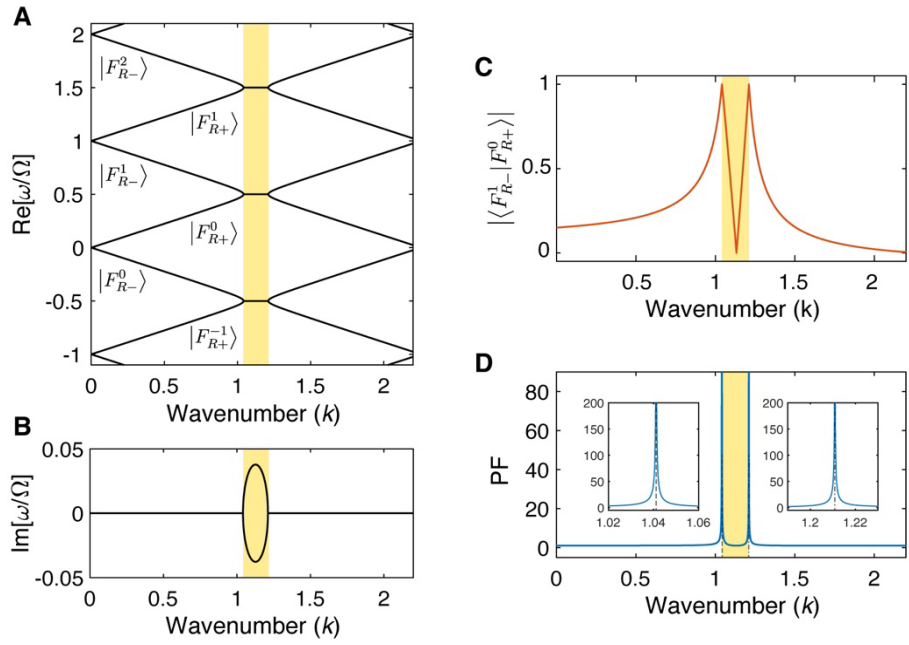

Fig. 1 Park *et al.*